\documentclass[aps,pra,superscriptaddress]{revtex4}
\usepackage{epsfig}
\usepackage{graphicx}
\usepackage{bm}
\usepackage{amssymb}
\newcommand{\zh}{\bm}

\newcommand{\zhr}{{\zh r}}

\newcommand{\zhp}{{\zh p}}

\newcommand{\Eq}[1]{Eq.\ (\ref{#1})}
\newcommand{\Eqs}[1]{Eqs.\ (\ref{#1})}

\begin{document}

\title{ Electron loss from hydrogen-like highly charged ions 
in collisions with electrons, protons and light atoms}

\author{ K.\ N.\ Lyashchenko }
 \affiliation{Department of Physics,
             St.\ Petersburg State University, 7/9 Universitetskaya nab.,
             St. Petersburg, 199034, Russia}
\author{O.\ Yu.\ Andreev}
\affiliation{Department of Physics,
             St.\ Petersburg State University, 7/9 Universitetskaya nab.,
             St. Petersburg, 199034, Russia}
\author{ A. B. Voitkiv }
\affiliation{ Institute for Theoretical Physics I,
Heinrich-Heine-University of D\"usseldorf, Germany }
\date{\today}

\begin{abstract}
We study electron loss from a hydrogen-like highly charged ion by the impact of equivelocity electrons and protons and also 
in collisions with hydrogen and helium. The collision velocity $v$ varies between $v_{min}$ and $v_{max}$, where  $v_{min}$ and $v_{max}$ 
correspond to the energy threshold $\varepsilon_{th}$ for electron loss in collisions with a free electron and to $\approx 5 \, \varepsilon_{th}$, respectively. 
Our results show that in this  range of $v$: i) compared to equivelocity electrons protons are more effective in inducing electron loss (due to a substantially larger volume of the effectively available final-state electron momentum space), 
ii) the relative (compared to protons) effectiveness of electron projectiles grows with increase in the atomic number of a highly charged ion, 
iii) a substantial part of the volume of the final-state-electron momentum space, kinematically available in collisions with electrons, is weaker populated in collisions with protons than with electrons, iv) even when the total loss cross sections in collisions with electrons and protons become already equal the spectra of the outgoing electrons still remain quite different in almost the entire volume of the final-state-electron momentum space. The points i) and iii), in particular, mean that in collisions with hydrogen target the contributions to the loss process from the interactions with the nucleus and the electron(s) of the atom would be to a large extent separated in the final-state-electron momentum space.      
\end{abstract}


\maketitle

\section{INTRODUCTION}
\par
For several decades fast ion-atom collisions, which occur at impact
velocities substantially exceeding the typical orbiting velocities of outer-shell
atomic electrons, have been a subject of extensive experimental and theoretical 
research. The exploration of different processes, accompanying such collisions,
are of great interest not only for the basic atomic physics research but
also have many applications in other fields of such as e.g. plasma physics,
astrophysics and radiation physics.

\par
With the advent of accelerators of relativistic heavy ions, in which the ions may 
reach velocities close to the speed of light $c$, much higher
impact energies and projectile charge states had become accessible for 
experiments on ion-atom collisions. The rich variety
of physical processes (e.g. target ionization and excitation; 
radiative and nonradiative electron capture;  
projectile-electron excitation and loss; 
free, bound-free and bound-bound pair production), 
which take place in ion-atom collisions at high impact energies, have
triggered great interest from the atomic physics community  
\cite{eichler-book,crothers-book,voitkiv2008-book}.

\par
Highly charged projectiles produced at accelerators of heavy ions 
often carry one or more bound electrons.
When such a projectile-ion collides with atomic or molecular targets 
these electrons can undergo transitions: 
the projectile can be excited or lose electron(s). 
In the rest frame of the ion this can be viewed as its excitation 
or "ionization" by the impact of the incident atom (or molecule). 
Electrons in a highly charged ion are very tightly bound. 
Therefore, in order to excite or remove them from the ion 
the momentum transfer between the ion and atom in the collision will 
in general be much larger than the momenta typical for the motion of 
outer-shall atomic electrons.

\par
Indeed, the minimum momentum transfer $q^{I}_{min}$ to the ion (given 
in the rest of the ion) and the minimum momentum transfer $q^{A}_{min}$ 
to the atom (given in the rest frame of the atom) read \cite{voitkiv2008-book}
\begin{eqnarray}
q^{I}_{min}
&=&\nonumber
\frac{\varepsilon_f - \varepsilon_i}{v}+ \frac{\epsilon_f - \epsilon_i}{\gamma v}\\
q^{A}_{min}
&=&\label{1u}
\frac{\varepsilon_f - \varepsilon_i}{v\gamma}+ \frac{\epsilon_f - \epsilon_i}{v}
\,.
\end{eqnarray}
Here, $\varepsilon_{i}$ ($\varepsilon_{f}$) is the energy of the initial (final) state 
of the electron of the ion given in the rest frame of the ion, $\epsilon_{i}$ ($\epsilon_{f}$) 
is the energy of the initial (final) state of the atom given in the rest frame of the atom. 
Further, $v$ is the collision velocity and $\gamma = 1/\sqrt{1-v^2/c^2}$ is the Lorentz factor of the collision. 
The energy difference 
$\varepsilon_f - \varepsilon_i$ is proportional to the square of the charge of the ion
reaching large values in case of very highly charged ions that makes $q^{A}_{min}$ 
quite big on the typical atomic scale. 

\par
For example, let us suppose that U$^{91+}(1s)$ loses an electron in the collision with an atom at an impact energy of  $1$ GeV/u ($v\approx120$ a.u., $\gamma\approx 2.07$). Applying \Eq{1u} to this case we obtain that $q^{I}_{min}$ and $q^{A}_{min}$ exceed $40$ a.u. and $19$ a.u., respectively. It is obvious that if the atom is very light then $q^{A}_{min}$ will be much larger than the typical orbiting momenta of {\it all} atomic electrons. It can be shown 
(see e.g. \cite{najjari2012}) that in such a case the cross section for excitation of or electron loss 
from the ion in collisions with an atom is approximated with very good accuracy by the following simple expression:  
\begin{eqnarray}
\sigma
&=& 
Z_A^2 \sigma_p + Z_A \sigma_e
\,, 
\label{atom} 
\end{eqnarray}
where $Z_A$ is the atomic number of the atom, and  $\sigma_p$ and $\sigma_e$ are 
the cross sections for the transition of an electron of the ion in collisions 
with equivelocity incident proton and electron \cite{foot-el-comp}, respectively. 
According to \Eq{atom} the process of electron loss in collisions with an atom 
reduces to two basic loss processes occurring in collisions with protons and electrons. 

\par
In the present paper we consider electron loss from a hydrogen-like highly charged ion (HCI) 
in collisions with equivelocity protons and electrons.
We also apply results obtained for collisions with electrons and protons 
in order to calculate electron loss cross sections in collisions with hydrogen and helium. 

We shall focus on the range of impact energies in which the collision velocity $v$ varies between $v_{min}$ and $v_{max}$, where  $v_{min}$ corresponds to the energy threshold of electron loss in collisions with a free electron and $v_{max}$ refers to the kinetic energy of the free electron $5$ times larger than the threshold value. This range has been chosen because of two main reasons. 
First, unlike very high-energy collisions, in which the energy of the incident electron greatly exceeds the threshold value and in which equivelocity electron and proton projectiles yield already practically the same loss cross sections, in the range of impact velocities of interest for the present study the differences between electron loss in collisions with these projectiles are expected to be quite large. 
Second, to our knowledge, for this range of impact velocities there have been 
no accurate calculations done for electron loss from HCIs in collisions 
with very light targets (like e.g. hydrogen and helium). Moreover, (accurate) experimental data 
for collisions of HCIs with such targets are also lacking in this range. 
Therefore, the results obtained in this paper 
can also serve as a guide for future experiments on electron loss. 

\par
There exists a number of experimental and theoretical data on the total cross section 
for electron loss from hydrogen-like HCIs in collisions with electrons \cite{deutsch1995,scofield1978,claytor1988,moores1995,marrs1994,marrs1997,rourke2001,watanabe2002}.
Besides, calculations are available for electron loss from such ions in collisions with 
various bare nuclei \cite{voitkiv2007a,voitkiv2007b} including protons.
However, to our knowledge, the processes of electron loss by electron and proton projectiles have never been considered within the scope of the same paper.  
(We note that a comparative study of excitation of HCIs by equivelocity protons and electrons was performed in \cite{najjari2013}.)
This, in particular, means that the important point about the relative effectiveness of these two types of projectiles in the process of electron loss remains largely unexplored. Besides, the lack of data on electron loss in collisions with equivelocity electrons and protons  makes it very difficult to apply results available for these projectiles to collisions with neutral atoms and molecules. 

\par
As was already mentioned, in this paper we intend also to consider electron loss from HCIs in collisions with hydrogen and helium. In this respect one should note that the electron loss in collisions with neutral atoms has been intensively studied (see, for example, \cite{dubois2004,watson2003,hulskotter1991,siegbert2014,siegbert2015,siegbert2016}, \cite{voitkiv2008-book}). However, for the range of impact energies which are of interest for present study a proper theory for this process in collisions with very light targets, where the interaction with electrons of the target yields a very substantial contribution to the total loss, has not yet been presented.

\par
The paper is organized as follows. In section II we describe two theoretical approaches. One of them is used for the consideration of electron loss from a HCI in collisions with electrons and the other one is applied to collisions with protons. Numerical results and discussion are presented in Section III. 
Section IV contains conclusions.

\par
The relativistic units ($\hbar = c = m_e = 1$)
are used throughout unless otherwise stated.

\section{GENERAL CONSIDERATION}

\subsection{Electron loss in collisions with an electron}

\par
Our description of electron loss from a hydrogen-like HCI by electron impact is based on the following main points.
First, the field of the nucleus of the HCI is so strong that the interaction of both electrons with this field should be taken into account to all orders. Therefore, in our description the Furry picture is used in which the interactions with the nucleus are fully taken into account from the onset. Second, since the interaction between the electrons is relatively very weak it is sufficient to take this interaction into account within the first order of the corresponding perturbation theory \cite{foot-wannier}.

\par
It is convenient to give the basic consideration of electron loss using the rest frame of the HCI. We assume that the nucleus of the HCI is infinitely heavy and take its position as the origin.

\par
Within the one-photon-exchange approximation the amplitude for electron loss reads
\begin{eqnarray}
U_{\mu_0, m_b,\mu_1,\mu_2}
&=&\label{ampl}
\alpha\int d^3 \zhr_1 d^3 \zhr_2 \, \bar{\psi}_{{\bf{p}}_1, \mu_1} (\zhr_1) \bar{\psi}_{{\bf{p}}_2, \mu_2}(\zhr_2)  \gamma^{\nu_1}\gamma^{\nu_2} I_{\nu_1 \nu_2}(|\varepsilon_1-\varepsilon_{0}|,|\zhr_1-\zhr_2|) \psi_{{\bf{p}}_0, \mu_0} (\zhr_1) \psi_{n_b j_b l_b,m_b} (\zhr_2)\\
&-&\nonumber
\alpha\int d^3 \zhr_1 d^3 \zhr_2 \, \bar{\psi}_{{\bf{p}}_1, \mu_1} (\zhr_1) \bar{\psi}_{{\bf{p}}_2, \mu_2}(\zhr_2)  \gamma^{\nu_1}\gamma^{\nu_2} I_{\nu_1 \nu_2}(|\varepsilon_1-\varepsilon_{n_b j_b}|,|\zhr_1-\zhr_2|) \psi_{n_b j_b l_b m_b} (\zhr_1) \psi_{{\bf{p}}_0, \mu_0} (\zhr_2)
\,.
\end{eqnarray}
Here, $\zhr_1$ and $\zhr_2$ are the electron coordinates, $\psi_{{\bf{p}}, \mu}$ is the wave function for an electron in the continuum with an asymptotical momentum  $\bf{p}$ (the corresponding energy is $\varepsilon=\sqrt{1+p^2}$) and polarization $\mu$.  The index $0$ in the continuum state refers to the incident electron and the indices $1$ and $2$ to the scattered and emitted electron. 
Further, $\psi_{n_b j_b l_b}$ represents the bound state wave function with the total energy $\varepsilon_{n_b j_b}$ where $n_b$, $j_b$, $l_b$ and $m_b$ are the principal quantum number, the total angular momentum, the orbital quantum number of the upper component and the projection of $j_b$, respectively.
In \Eq{ampl} $\gamma^{\nu_i}$ are the gamma matrices ($i=1,2$), $\alpha$ 
is fine-structure constant and $I_{\nu_1 \nu_2}$ denotes the photon propagator. 
In the Feynman gauge the propagator is given by (see e.g. \cite{andreev2008})

\begin{eqnarray}
I_{\nu_1 \nu_2} \left(|\varepsilon-\varepsilon_{n_b j_b}|,|{\bf {\zhr}_1}-\zhr_2| \right)
&=&
\frac{g_{\nu_1\nu_2}\exp{[i|\varepsilon-\varepsilon_{n_b j_b}||\zhr_1-\zhr_2|]}}{|\zhr_1-\zhr_2|}
\,,
\end{eqnarray}
where $g_{\nu_1\nu_2}$ is the metric tensor ($g_{11}=-g_{22}=-g_{33}=-g_{44}=1$ and $g_{\nu_1,\nu_2}=0$ if $\nu_1 \neq \nu_2$).

\par
Wave functions of the electrons from the continuum spectra $\psi_{{\bf{p}}, \mu}$ are described
by the following expansion \cite{akhiezer1965}
\begin{eqnarray}
\psi_{\zhp, \mu}(\zhr)
&=&\label{psie}
\int d\epsilon \sum_{jlm} a_{{\zhp} \mu, \epsilon jlm}\psi_{\epsilon jlm}(\zhr)
\,,
\end{eqnarray}
where $\psi_{\epsilon jlm}$ is a solution of the Dirac equation with a field of the HSI's nucleus and the coefficients $a_{{\zhp} \mu, \epsilon jlm}$ have a following form
\begin{eqnarray}
a_{{\zhp} \mu, \epsilon jlm}
&=&\label{coefa}
\frac{(2\pi)^{3/2}}{\sqrt{p \epsilon}}i^l e^{\pm i\phi_{jl}}
(\Omega^+_{jlm}(\zhp), \upsilon_{\mu}(\zhp))
\delta(\epsilon-\varepsilon)
\,.
\end{eqnarray}
In \Eq{coefa} the $\phi_{jl}$ ($-\phi_{jl}$) is the Coulomb phase shift for the incident (emitted or scattered) electron,
$\Omega_{jlm}(\zhp)$ is the spherical spinor (tensor spherical harmonic \cite{varshalovich1988lkn})
and $\upsilon_{\mu}(\zhp)$ is the spinor with
projection $\mu=\pm 1/2$ on the electron momentum ($\zhp$)
\begin{eqnarray}
\frac{\zhp \, \hat{\zh\sigma}}{2p} \upsilon_{\mu}(\zhp)
&=&
\mu
\upsilon_{\mu}(\zhp)
\,,
\end{eqnarray}
where
$\hat{\zh\sigma}$
is the vector consisting of the Pauli matrices.

\par
Assuming that the incident electron is unpolarized and polarizations 
of the scattered and emitted electrons are not detected, 
the fully differential cross section for electron loss is given by
\begin{eqnarray}
\frac{d\sigma}{d\varepsilon_{1} d\Omega_{p_1}d\varepsilon_{2} d\Omega_{p_2}}
&=&\label{sect_1}
\frac{\pi}{2}\sum_{\mu_0, m_b, \mu_1,\mu_2} |U_{\mu_0,m_b,\mu_1,\mu_2}|^2 \frac{\varepsilon_{0}}{p_0} \frac{\varepsilon_{1}p_1 \varepsilon_{2}p_2}{(2\pi)^6}\delta(\varepsilon_{1s}+\varepsilon_{0}-\varepsilon_{1}-\varepsilon_{2})
\,,
\end{eqnarray}
where $\Omega_{p_1}$ and $\Omega_{p_2}$ are momentum solid angles of the emitted and scattered electrons.

In the present paper we shall not discuss the above cross section (\ref{sect_1}) but instead in section III we consider the cross section differential in energy and solid angle of one of the outgoing electrons 
which is given by   
\begin{eqnarray}
\frac{d\sigma}{d\varepsilon d\Omega_{p}} 
&=&\label{sect_1a}
\int d\varepsilon_{2} d\Omega_{p_2} \frac{d\sigma}{d\varepsilon d\Omega_{p}d\varepsilon_{2} d\Omega_{p_2}}  
\\  
&=&\nonumber
\int d\varepsilon_{1} d\Omega_{p_1} \frac{d\sigma}{d\varepsilon_{1} d\Omega_{p_1}d\varepsilon d\Omega_{p}} 
\, . 
\end{eqnarray}

\subsection{ Electron loss in collisions with protons } 

Let us now consider electron loss from a hydrogen-like HCI in collisions with a proton. Like in the previous subsection, we shall give the basic consideration of this process using the rest frame of the HCI and take the position of its nucleus as the origin. In our description we shall regard the proton as a Dirac particle. Since the interaction between the electron and the incident particle is again comparatively very weak, the process of electron loss can be treated within the first order of perturbation theory in this interaction \cite{foot-proton-cusp}. 
Then the transition amplitude for electron loss by proton impact is given by 
\begin{eqnarray}
S_{\mu_{i}, m_b,\mu_{f},\mu}
&=&\label{trans_ampl_proton}
\int d^3{\bf {R}} \, d^3 \zhr \,  \bar{\Phi}^{(-)}_{{\bf{P}}_{f}, \eta_{f}} ({\bf {R}}) \bar{\psi}_{{\bf{p}}, \mu}(\zhr)  (-\alpha) \gamma^{\nu_1}\gamma^{\nu_2} I_{\nu_1 \nu_2} \left(|\varepsilon-\varepsilon_{n_b j_b}|,|\zhr-{\bf {R}}| \right) \Phi^{(+)}_{{\bf{P}}_{i}, \eta_{i}} ({\bf {R}}) \psi_{n_b j_b l_b m_b} (\zhr)
\,.
\end{eqnarray}
Here, $\zhr$ and ${\bf {R}}$ are coordinates of the electron and proton, respectively, $\Phi^{(+)}_{{\bf{P}}_{i}, \eta_{i}}$ ($\Phi^{(-)}_{{\bf{P}}_{f}, \eta_{f}}$) is the wave function for the incident (scattered) proton with an asymptotical momentum  $\bf{P}_{i}$ ($\bf{P}_{f}$), the total energy $E_{i}$ ($E_{f}$) and polarization $\eta_{i}$ ($\eta_{f}$). The other notations in \Eq{trans_ampl_proton} are the same as in \Eq{ampl}.

\par
The interaction between the electron and the nucleus of the HCI is very strong and should be fully taken into account. This is done in our treatment by describing states of the bound ($\psi_{n_b j_b l_b m_b}$) and emitted ($\psi_{{\bf{p}}, \mu}$) electron in 
\Eq{trans_ampl_proton} using the Coulomb-Dirac states.

\par
The strength of the interaction between the proton and the nucleus of the HCI is characterized by the Sommerfeld parameter $\alpha \frac{Z_I}{v}$, where $Z_I$ denotes the atomic number of the HCI and $v$ is the asymptotic velocity of the incident proton. 
Because $Z_{I}$ can be very large and $v$ can not exceed the speed of light 
this parameter in our case may be not much smaller than $1$.
This means that the interaction between the proton and the nucleus of the HCI may not be simply disregarded. Instead, it should be taken into account to all orders. However, the description of the proton in our process can be drastically simplified if we remark that, because of its huge (compared to the electron) mass, it has an enormous momentum. As a result, the Coulomb-Dirac states for the incident ($\Phi^{(+)}_{{\bf{P}}_{i}, \eta_{i}}$) and scattered ($\Phi^{(-)}_{{\bf{P}}_{f}, \eta_{f}}$) proton can be taken in the eikonal approximation in which they read (see, for instance, \cite{landau_v4})
\begin{eqnarray}
\Phi^{(+)}_{{\bf{P}}_{i}, \eta_{i}} ({\bf {R}})
&\approx&\label{wfp_1}
\frac{1}{\sqrt{2E_{i}}} \exp\left[ i\frac{\alpha E_{i}}{P_i}\ln(P_{i}R-{\bf{P}}_{i} 
\cdot {\bf{R}}) \right]
\exp\left[ i {\bf{P}}_{i} \cdot {\bf{R}} \right] u_{P_i \eta_i}
\,,
\end{eqnarray}

\begin{eqnarray}
\Phi^{(-)}_{{\bf{P}}_{f}, \eta_{f}} ({\bf {R}})
&\approx&\label{wfp_2}
\frac{1}{\sqrt{2E_{f}}} \exp\left[ -i\frac{\alpha E_{f}}{P_f}\ln(P_{f}R+{\bf{P}}_{f} \cdot {\bf{R}}) \right]
\exp\left[ i {\bf{P}}_{f} \cdot {\bf{R}} \right] u_{P_f \eta_f}
\,,
\end{eqnarray}
where $u_{P_i \eta_i}$ and $u_{P_f \eta_f}$ are constant bispinor amplitudes for the corresponding plane waves.

\par
The next key point in our description of the proton is to remark that the changes in the proton momentum 
and energy in the collision are negligibly small compared to their corresponding initial values. Taking this into account enables us to perform analytically the integration of the transition probability over the momenta of the scattered proton and then sum and average the result over polarization of the final and initial states of the proton. By performing all these steps we obtain the following expression for the differential cross section of electron loss 
\begin{eqnarray}
\frac{d\sigma}{d\varepsilon d\Omega}
&=&\label{sect_2}
\frac{\varepsilon p}{(2 \pi)^3} \frac{2\alpha^2}{v^2}
\sum_{\mu m_b}
\int  d^2 {\bf{q}}_{\bot} \,  \left| \frac{\int d^3 \zhr \,\bar{\psi}_{{\bf{p}}, \mu}(\zhr) \exp{(i{\bf{q}} \cdot \zhr)} (\gamma_0 -v\gamma_3) \psi_{n_b j_b l_b m_b} (\zhr) }{q'^2}
\right|^2
\,,
\end{eqnarray}
where $\Omega$ is the momentum solid angle of the emitted electron  
and ${\bf q}$ is the change in the momentum of the proton    
\begin{eqnarray}
{\bf{q}}
&=&
({\bf{q}}_{\bot}, q_{min}); \phantom{12} 
q_{min}=\frac{\varepsilon-\varepsilon_{n_b j_b}}{v} 
\,, 
\end{eqnarray}
representing the momentum transfer to the HCI. 
The quantity ${\bf q}'$, which also enters Eq. (\ref{sect_2}) and which is given by 
\begin{eqnarray}
{\bf{q}}'=
\left( {\bf{q}}_{\bot}, \frac{q_{min}}{\gamma} \right),  
\end{eqnarray}
has the meaning of the change in the momentum of the proton 
in the reference frame where the proton is initially at rest. 

The integration in (\ref{sect_2}) runs over the two-dimensional vector 
of the transverse momentum transfer ${\bf{q}}_{\bot}$.  
On the scale of the electron momentum the upper limit of the integration 
over the absolute value of  ${\bf{q}}_{\bot}$ in (\ref{sect_2}) 
can be safely set to infinity.  

To conclude this subsection we note that the expression (\ref{sect_2}) 
can also be derived within the so called semi-classical approximation 
in which the incident proton 
is regarded as a classical particle moving along a straight-line trajectory. 
It is remarkable 
that our more general treatment, in which the proton is described 
quantum-mechanically and in which its interaction with the nucleus 
of the HCI is fully taken into account, yields the same result 
for the cross section (\ref{sect_2}) giving, thus, 
a direct proof of the validity of the semiclassical approximation in our case.      

\subsection{ Limitations of the approaches presented in subsections A and B} 

In the approaches, presented in the above subsections, the interaction between 
the incident particle (an electron, a proton) with the electron of the HCI 
is treated within the first order of perturbation theory. For collisions with HCIs 
this is in general an excellent approximation. However, it nevertheless 
breaks down when the emitted electron and the scattered particle 
have very close velocities. 

In case of electron loss by proton impact  
such a situation would correspond to the so called electron capture into 
the projectile continuum and it can be described by using e.g. distorted wave models  
(see \cite{voitkiv2007b}). Since the momentum space available for the emitted 
electron is very large the capture to the projectile continuum 
affects the electron emission pattern 
only in a very small part of this space and 
has practically no influence on 
the total loss cross section.  

If electron loss occurs in collisions with electrons 
and in the final state both electrons have very close 
velocities then the electron-electron repulsion can influence 
the shape of the electron spectra. In electron loss from HCIs 
by electron impact at collision energies sufficiently far 
from the threshold value this influence is of importance 
just for a small part of the final electron momentum space having, therefore, 
a very limited overall effect on the spectral shape and very weak impact on 
the total cross section. The simplest way to account qualitatively 
for the electron-electron repulsion in this case is to introduce the so called Gamov factor, 
$G = 2 \pi \eta/(\exp(2 \pi \eta) -1 )$ with $\eta = e^2/\hbar \Delta v_e$, 
where $e$, $\hbar$ and $\Delta v_e$ is the electron charge, the Planck constant and 
the absolute value of the difference in the electron velocities.  
    
If electron loss by electron impact takes place at impact energies very close 
to the threshold then both the electron spectra and the total cross section may be 
strongly influenced by the electron-electron repulsion. It is obvious that the approach 
described in subsection A becomes inaccurate in this (so called Wannier \cite{wannier}) regime 
of electron loss.  
    
\section{ Results and discussion }

\par
Let us now consider results of our calculations for the cross sections obtained 
by using \Eqs{sect_1a}, (\ref{sect_2}) and (\ref{atom}). 

\subsection{Electron energy-angular distributions} 

We start with the cross sections $\frac{d\sigma}{d\varepsilon d\Omega}$, where $\varepsilon$ and $\Omega$ are the kinetic energy and the momentum solid angle, respectively, of the electron in the final state. This cross section represents the energy-angular distribution 
for the outgoing electron, it is independent of the azimuthal emission  angle.  
We note that in case of collisions with an electron 
the quantities $\varepsilon$ and $\Omega$ refer to any of the two electrons in the final state (see Eq. \ref{sect_1a}).

One should note that, while the total cross section for loss from HCIs 
by electron impact was calculated \cite{foot1} in the past 
\cite{deutsch1995,scofield1978,moores1995,rourke2001},     
and the fully differential cross section had been explored 
for a close problem of ionization of the $K$-shell of heavy atoms by electron impact 
(see for review \cite{nakel1999}),   
we are not aware about any previous calculations of the cross section 
$\frac{d\sigma}{d\varepsilon d\Omega}$ in case of electron projectiles.   

\begin{figure}[h]
\begin{minipage}{40pc}
\includegraphics[width=40pc]{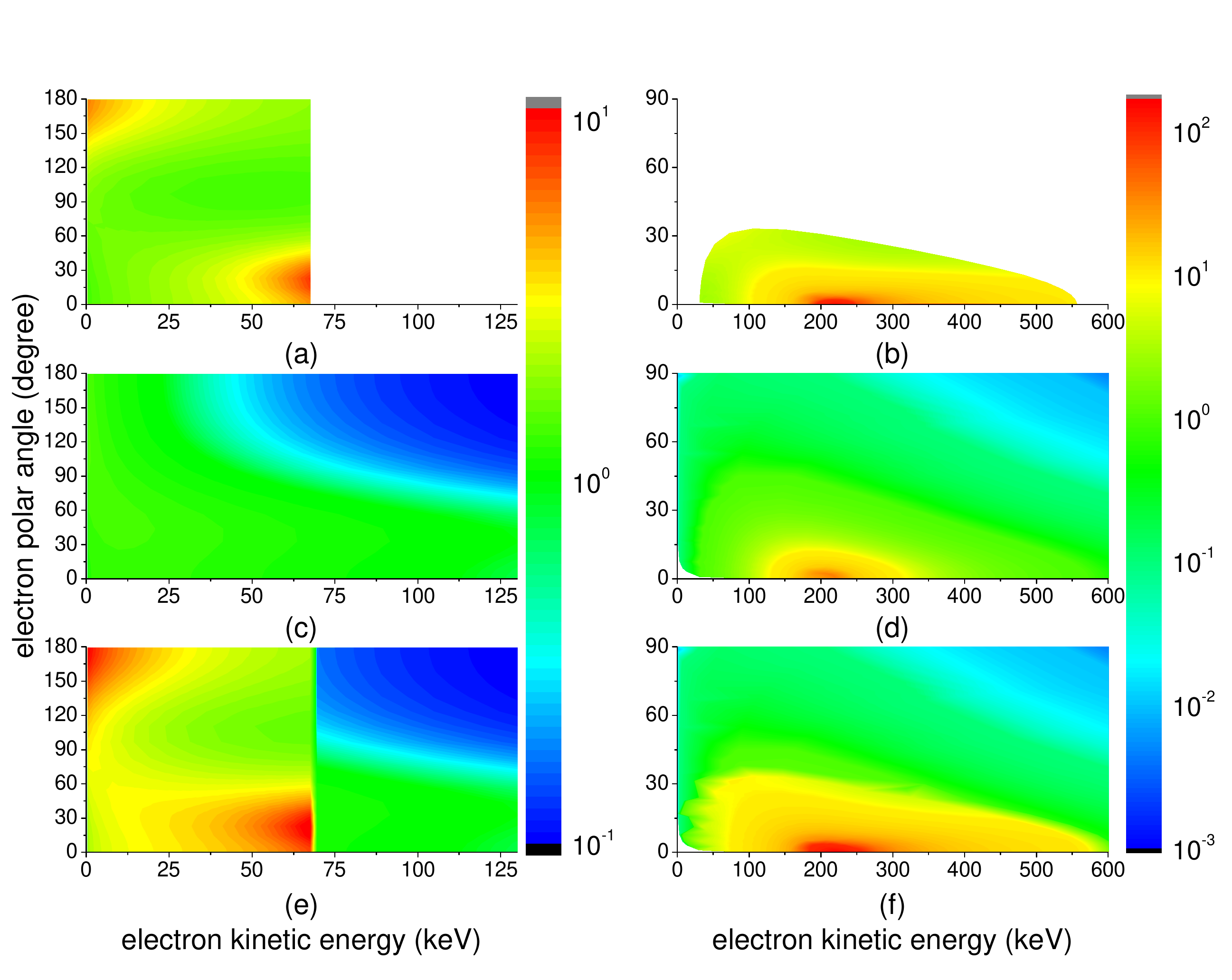}
\caption{The doubly differential cross sections (in b/MeV/sr) in collisions of U$^{91+}$(1s) with equivelocity electrons, protons and hydrogen. In the rest frame of the HCI the polar angle for the outgoing electron is counted from the direction of the motion of the incident particle whereas in 
the laboratory frame this angle is counted from the direction of the HCI's motion. 
The left panel shows results in the rest frame of uranium for 
incident $200$ keV electrons (a), $367.2$ MeV protons (c) 
and $367.4$ MeV hydrogen (e). The right panel presents results for the laboratory frame 
in which $364.546$ MeV/u U$^{91+}$(1s) collides with electrons (b), protons (d) and hydrogen (f). }
\label{fig1}
\end{minipage}\hspace{2pc}%
\end{figure}

\begin{figure}[h]
\begin{minipage}{40pc}
\includegraphics[width=40pc]{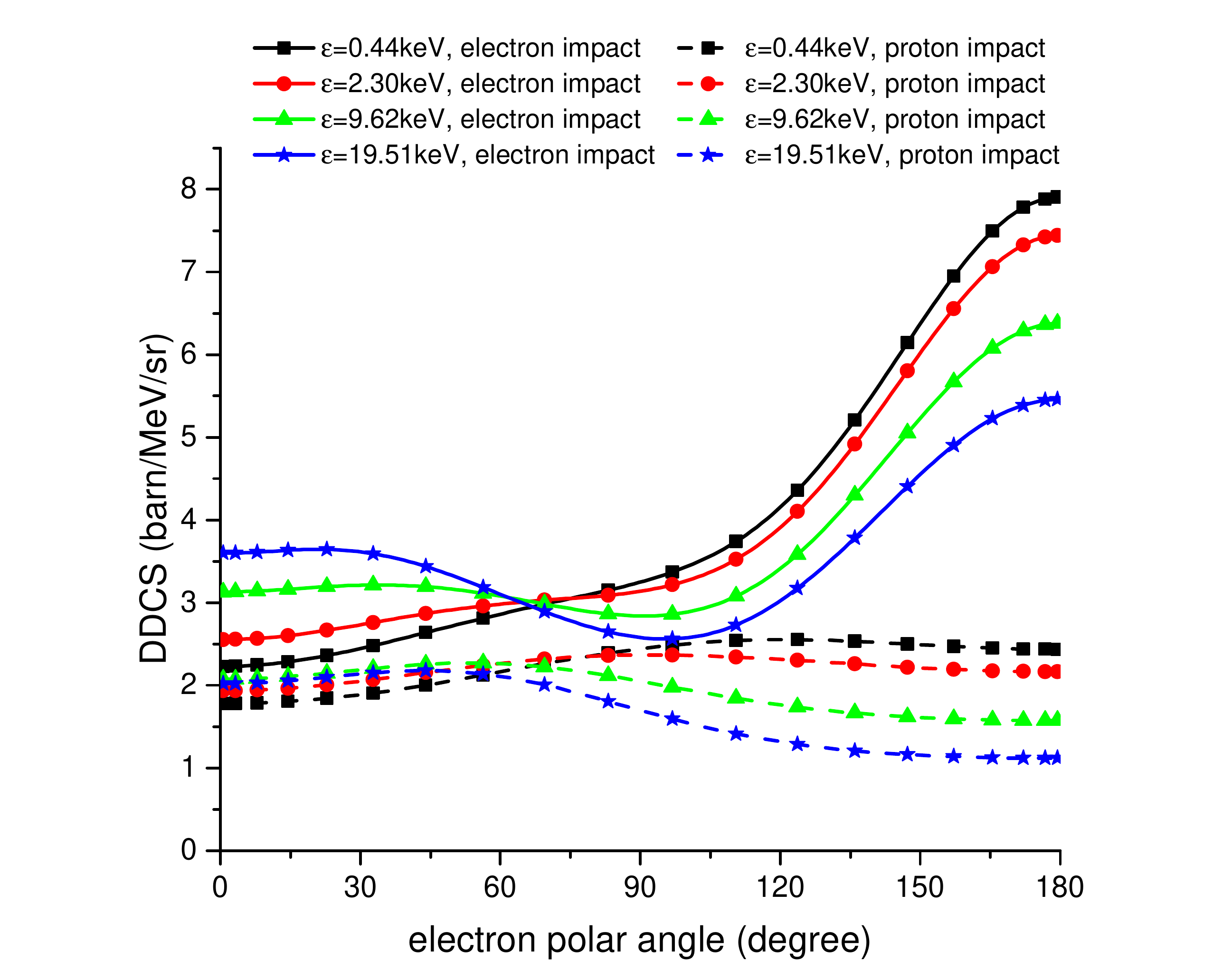}
\caption{The doubly differential cross sections in collisions of U$^{91+}$ with $200$ keV electrons and equivelocity protons given in the rest frame of the HCI as a function of the polar angle of the outgoing electron at a fixed electron energy. 
The angle is counted from the direction of the motion of the incident particle. }
\label{fig1-add}
\end{minipage}\hspace{2pc}%
\end{figure}

\par 
In Fig. \ref{fig1} we present the energy-angular distributions 
of outgoing electrons in collisions of U$^{91+}(1s)$ with $200$ keV electrons and with equivelocity protons having an impact energy of $367.2$ MeV. Besides, in this figure we show also results for collisions with atomic hydrogen (in case of collisions with randomly oriented hydrogen molecules the results for atomic hydrogen should simply be multiplied by $2$). The latter were obtained by summing, according to \Eq{atom}, the corresponding results for collisions with equivelocity electrons and protons \cite{foot-compton-profile}.  The energy-angular distributions were calculated for two reference frames: the rest frame of the HCI and the frame in which the incident electron (proton, atom) is initially at rest. 

\par
The process of electron loss looks simpler when it is considered 
in the rest frame of the HCI. 
The general observation which follows from the results for 
the energy-angular distributions 
in this frame (see the left panel of Fig. \ref{fig1}) is that  
the shapes of the electron spectra 
in collisions with electrons and protons are very different. 

In particular, in collisions 
with protons the spectrum of electrons is more extended 
to higher energies compared to collisions with electrons. 
This is quite expected since compared to an electron 
an equivelocity proton carries much more energy 
and, therefore, the kinematically available volume in 
the final-state-electron momentum space 
is orders of magnitude larger in collisions with protons. 
Although, due to the huge difference 
in masses and the absence of very high frequency 
components in the field of the proton, 
just a tiny part of this volume can be noticeably populated,  
the resulting effective volume in collisions with protons 
still substantially exceeds that which is available 
in collisions with electrons. 

The results presented in Fig. \ref{fig1} also show 
that a considerable part of the momentum space 
in the final state, which is kinematically available 
in collisions with electrons, 
is substantially weaker populated in collisions with protons than 
in those with equivelocity electrons. 

From the above two observations it, in particular, follows 
that in collisions with atomic (or molecular) hydrogen 
the contributions to the loss process from the interactions 
with the nucleus (nuclei)  
and the electron(s) of the atom (molecule) 
are to a large extent separated 
in the final-state-electron momentum space.     

We also observe in Figs. \ref{fig1} - \ref{fig1-add}  
that for a given energy of the outgoing electron  
the shape of the angular distributions 
in collisions with electrons and protons 
may differ very substantially. In particular, 
a strong increase at large angles can be traced back 
to originate due to constructive interference 
between the direct and exchange contributions 
to the transition amplitude (\ref{ampl}).   
  
\begin{figure}[h]
\begin{minipage}{40pc}
\includegraphics[width=40pc]{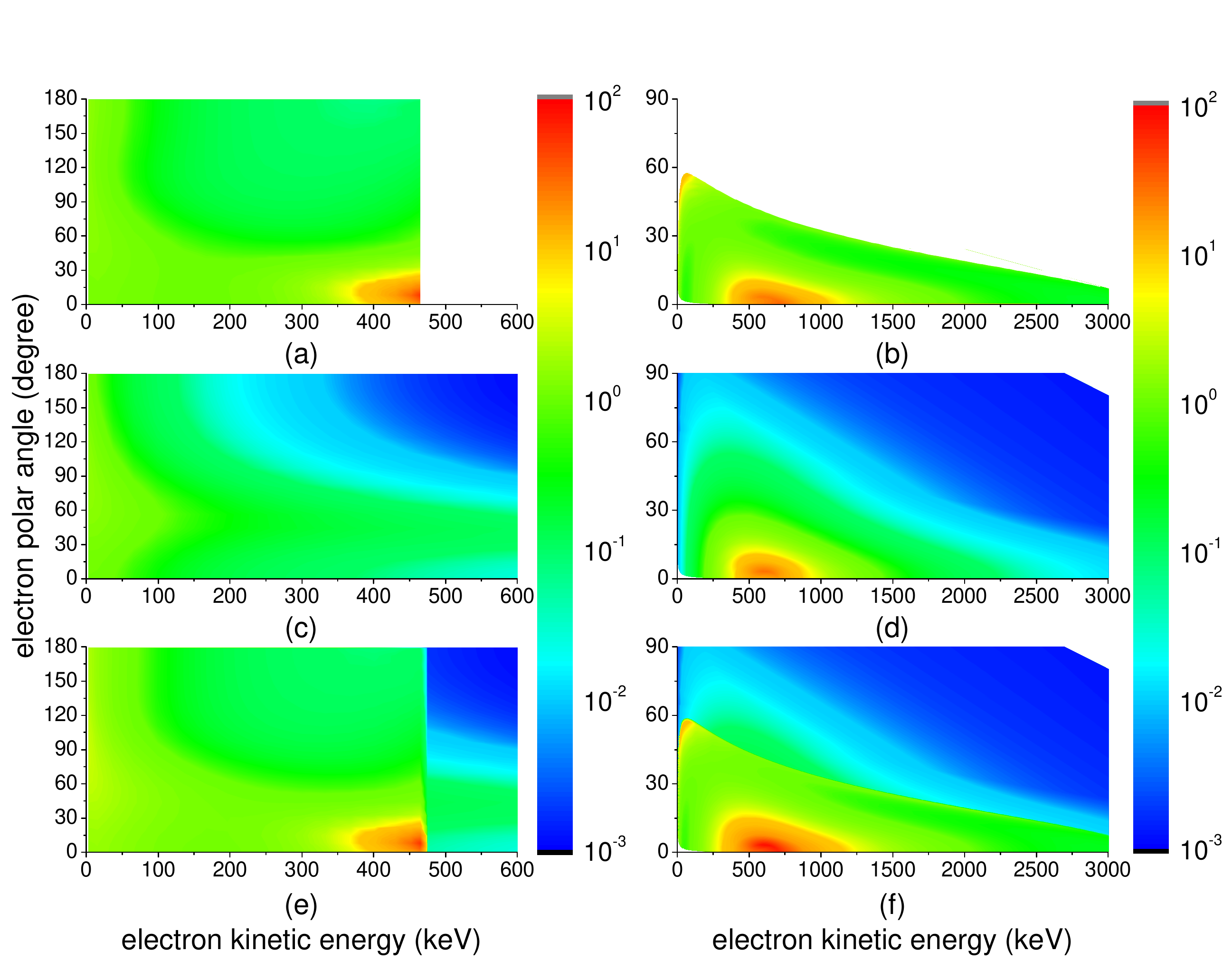} 
\caption{ Same as in Fig. \ref{fig1} but 
for $600$ keV electrons, 
$1.1$ GeV protons, $1.1$ GeV hydrogen and $1.09$ GeV/u U$^{91+}$(1s). } 
\label{fig2}
\end{minipage}\hspace{2pc}%
\end{figure}

\begin{figure}[h]
\begin{minipage}{40pc}
\includegraphics[width=40pc]{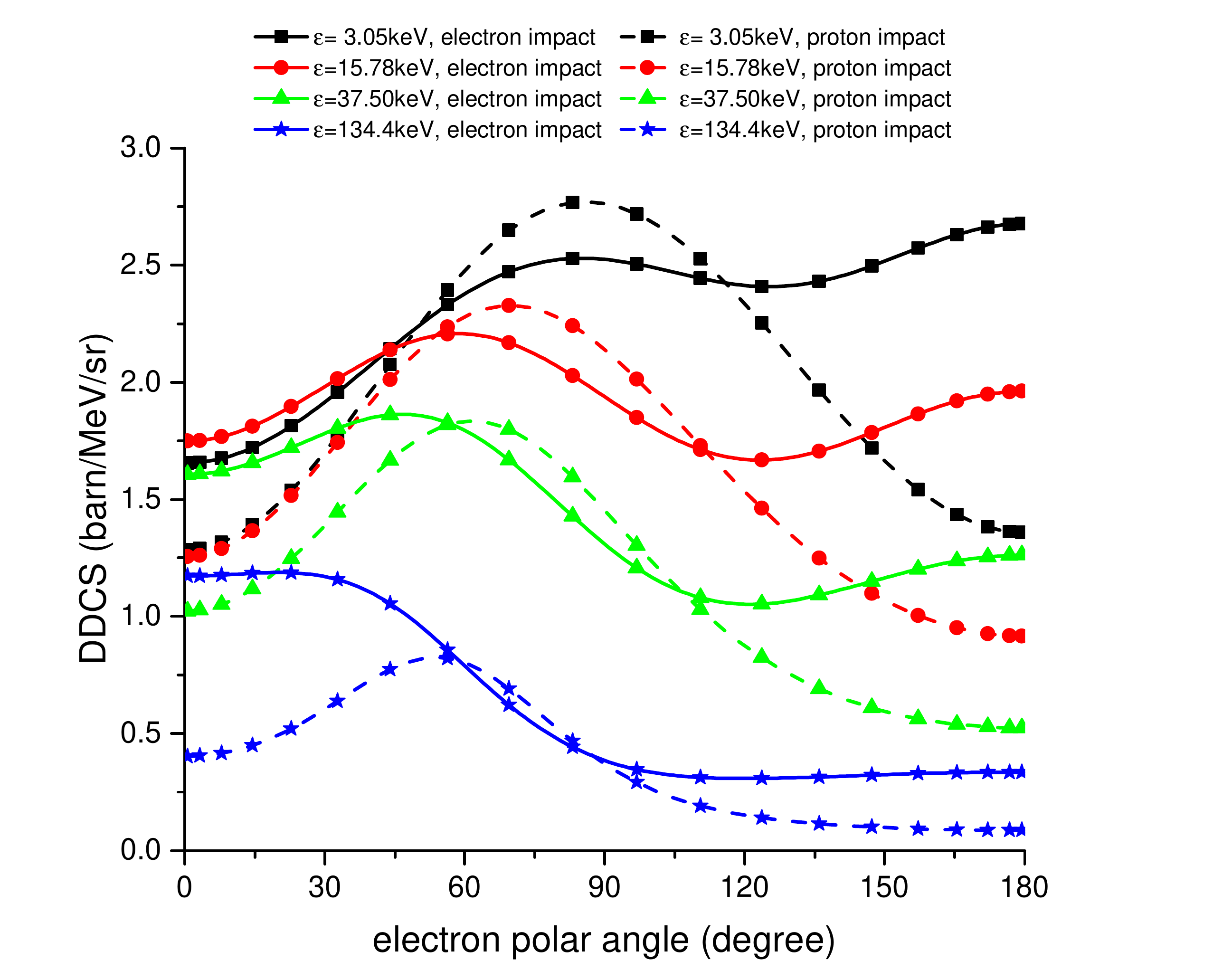}
\caption{ Same as in Fig. \ref{fig1-add} but 
for incident $600$ keV electrons 
and equivelocity protons. }
\label{fig2-add}
\end{minipage}\hspace{2pc}%
\end{figure}

\par
In Figs. \ref{fig2} and \ref{fig2-add} 
we show results for the energy-angular distributions 
of the outgoing electrons in collisions of U$^{91+}(1s)$ 
with $600$ keV electrons 
and equivelocity protons with the corresponding impact energy 
of $1.1$ GeV as well as with $1.1$ GeV atomic hydrogen. 
Like in the previous case we observe in Fig. \ref{fig2} 
that, compared to collisions with electrons, 
in collisions with equivelocity protons 
the spectra extend to higher energies. 
Besides, in the momentum space,  
kinematically available in collisions with electrons, 
there are parts which are substantially stronger populated 
in collisions with electrons than with 
equivelocity protons. 

There are also noticeable differences in the shape of electron spectra 
presented in Figs. \ref{fig1}-\ref{fig1-add} and \ref{fig2}-\ref{fig2-add} 
which are especially visible in case of electron loss by electron impact. 
For instance, in collisions with $200$ keV electrons 
the angular distribution of low-energy electrons is quite asymmetric with a pronounced 
maximum at $180^\circ$ whereas at $600$ keV impact energy the distribution 
of these electrons is more homogeneous. Also the separation between the maxima 
at low and high energies becomes better with increasing the impact energy reflecting 
a better separation between the electrons in the phase space that enables one 
to almost unambiguously identify the maximum at lower and higher energies with the emitted and 
scattered electron, respectively.        

\begin{figure}[h]
\begin{minipage}{40pc}
\includegraphics[width=40pc]{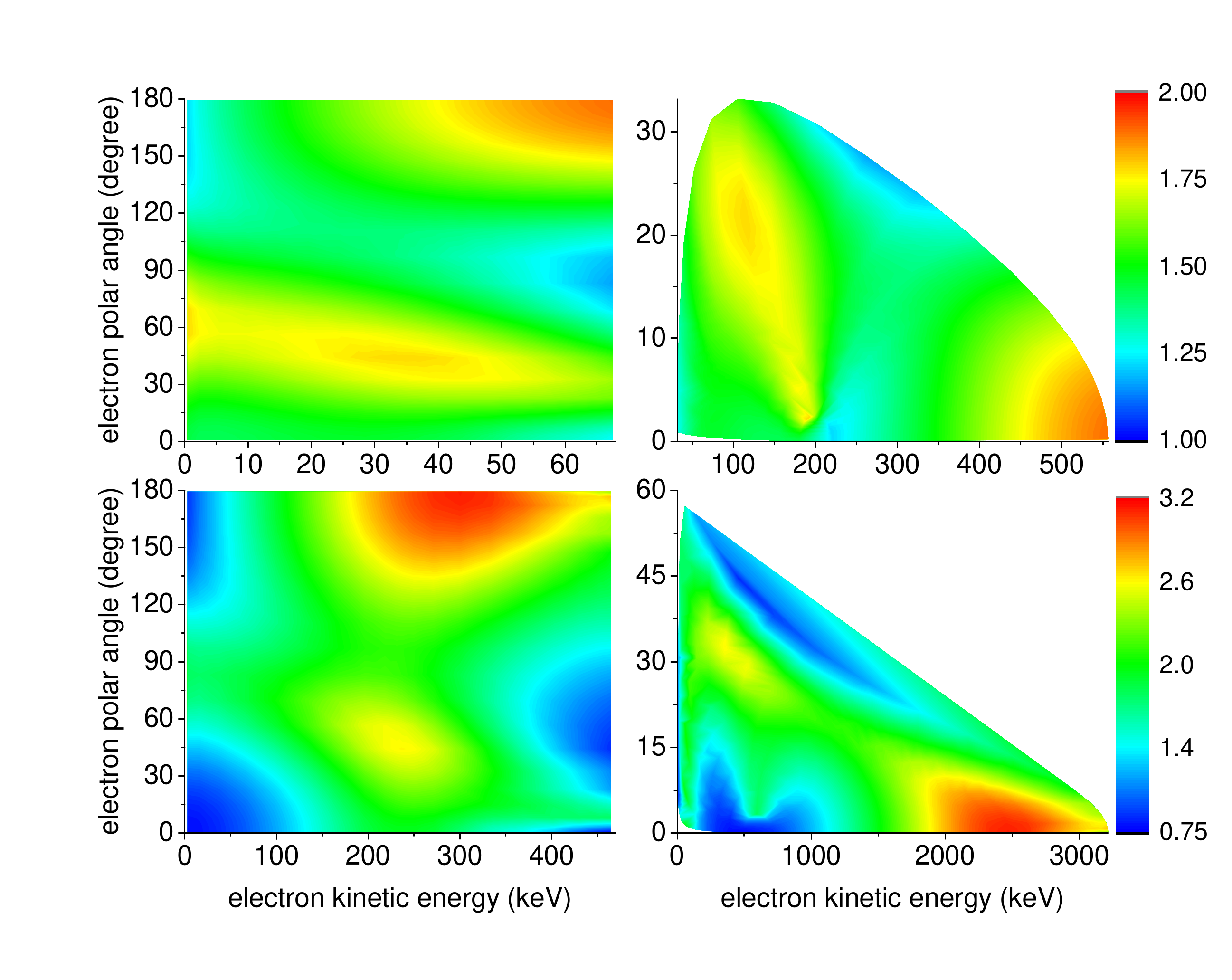}
\caption{The ratio 
$\frac{d\sigma}{d\varepsilon d\Omega}/\frac{d\sigma^{C}}{d\varepsilon d\Omega}$ 
for electron loss from U$^{91+}$(1s) in collisions with $200$ keV (the upper part) and $600$ keV 
(the lower part) electrons, where $\frac{d\sigma^{C}}{d\varepsilon d\Omega}$ was calculated by ignoring 
the Breit interaction. The left and right panels show the ratio in the rest frame of the HCI and 
the frame where the incident electron is initially at rest, respectively. 
The polar angle of the electron is counted as in Figs. (\ref{fig1}) and (\ref{fig2}). }
\label{fig3}
\end{minipage}\hspace{2pc}%
\end{figure}

The interaction between the electrons can be considered as the sum of 
the (unretarded) Coulomb and the Breit interactions. The latter represents 
the (main) relativistic correction to the former and becomes of importance 
when the energy of the incident electron and/or the atomic of the HCI 
are/is sufficiently high. 
In Fig.  \ref{fig3} we show the ratio 
$\frac{d\sigma}{d\varepsilon d\Omega}/\frac{d\sigma^{C}}{d\varepsilon d\Omega}$  
for electron loss from U$^{91+}$(1s) by the impact of $200$ and $600$ keV electrons, 
where the cross section $\frac{d\sigma^{C}}{d\varepsilon d\Omega}$ was calculated by ignoring 
the Breit interaction. It follows from the figure that the Breit interaction very substantially 
influences the energy-angular distribution. Besides, 
the results shown in Fig. \ref{fig3} quite clearly suggest that 
this interaction increases the total number of electron loss events. 

\begin{figure}[h]
\begin{minipage}{40pc}
\includegraphics[width=40pc]{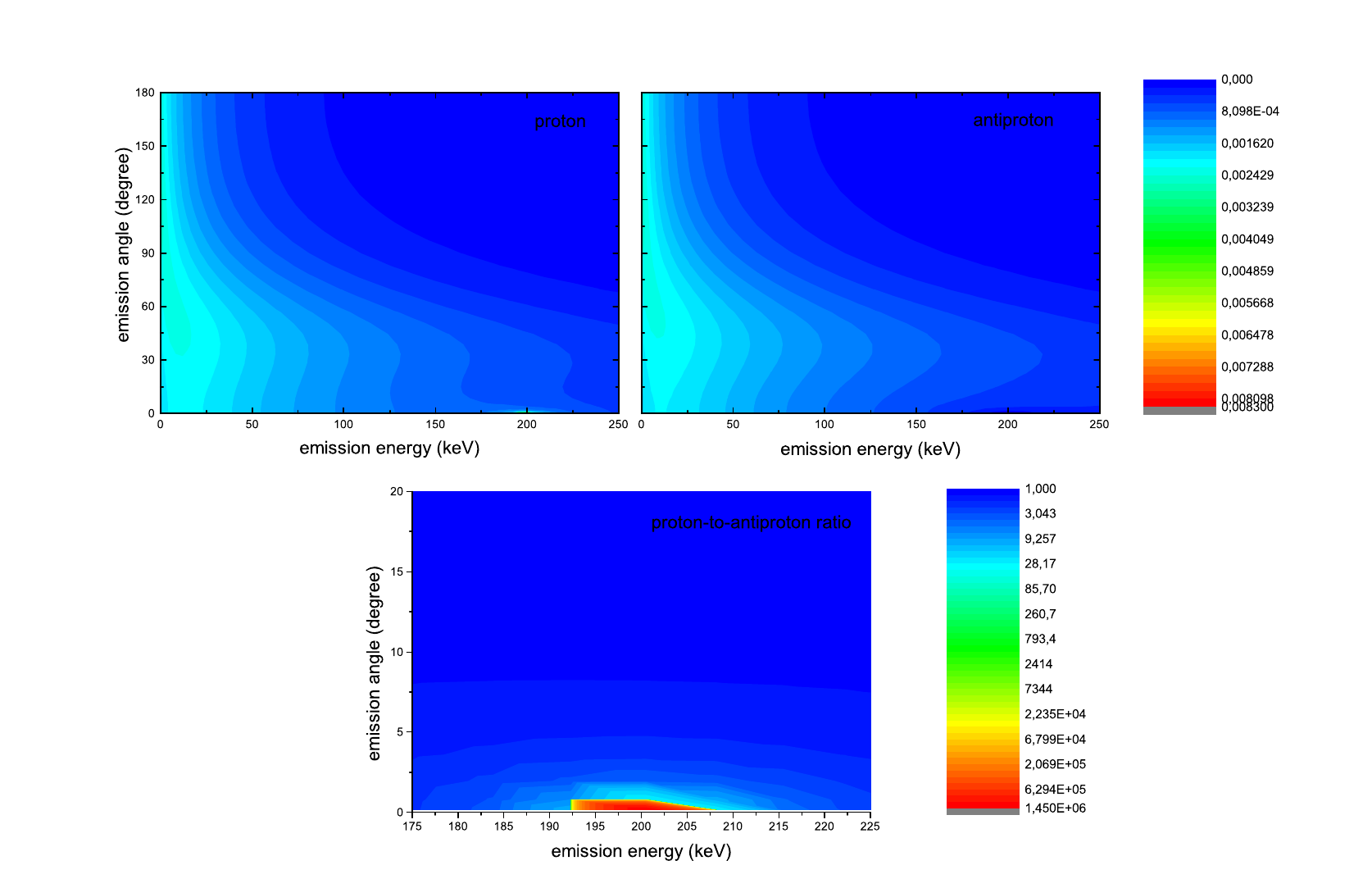}
\caption{The cross section $\frac{d\sigma}{d\varepsilon d\Omega}$ 
(in b/keV/sr) for electron loss from U$^{91+}$(1s) in collisions 
with $367.2$ MeV protons (the upper left part) and antiprotons 
(the upper right part). The lower panel represents 
the proton-to-antiproton cross section ratio.}
\label{fig-pr-apr}
\end{minipage}\hspace{2pc}%
\end{figure}

\vspace{0.25cm} 

We conclude this subsection by considering the relative importance of the restrictions 
of our approach in case of collisions with protons which were discussed in subsection II.C. 
We illustrate this importance in figure \ref{fig-pr-apr} where results of calculations 
for electron emission from U$^{91+}$(1s) in collisions with $367.2$ MeV protons 
and equivelocity antiprotons are presented. These results were obtained by using 
a distorted-wave approach of \cite{voitkiv2007b} in which the interaction 
between the emitted electron and the projectile is taken into account 
(for more detail of the approach see \cite{voitkiv2007b}).    
It follows from the figure that for the overwhelming part 
of the momentum space of the emitted electron the differences between 
collisions with protons and antiprotons are negligible and that only 
when the electron velocity is close to that of the projectile the emission patterns 
begin to differ drastically.   
 
\subsection{Total cross section for electron loss}

\begin{figure}[h]
\begin{minipage}{40pc}
\includegraphics[width=40pc]{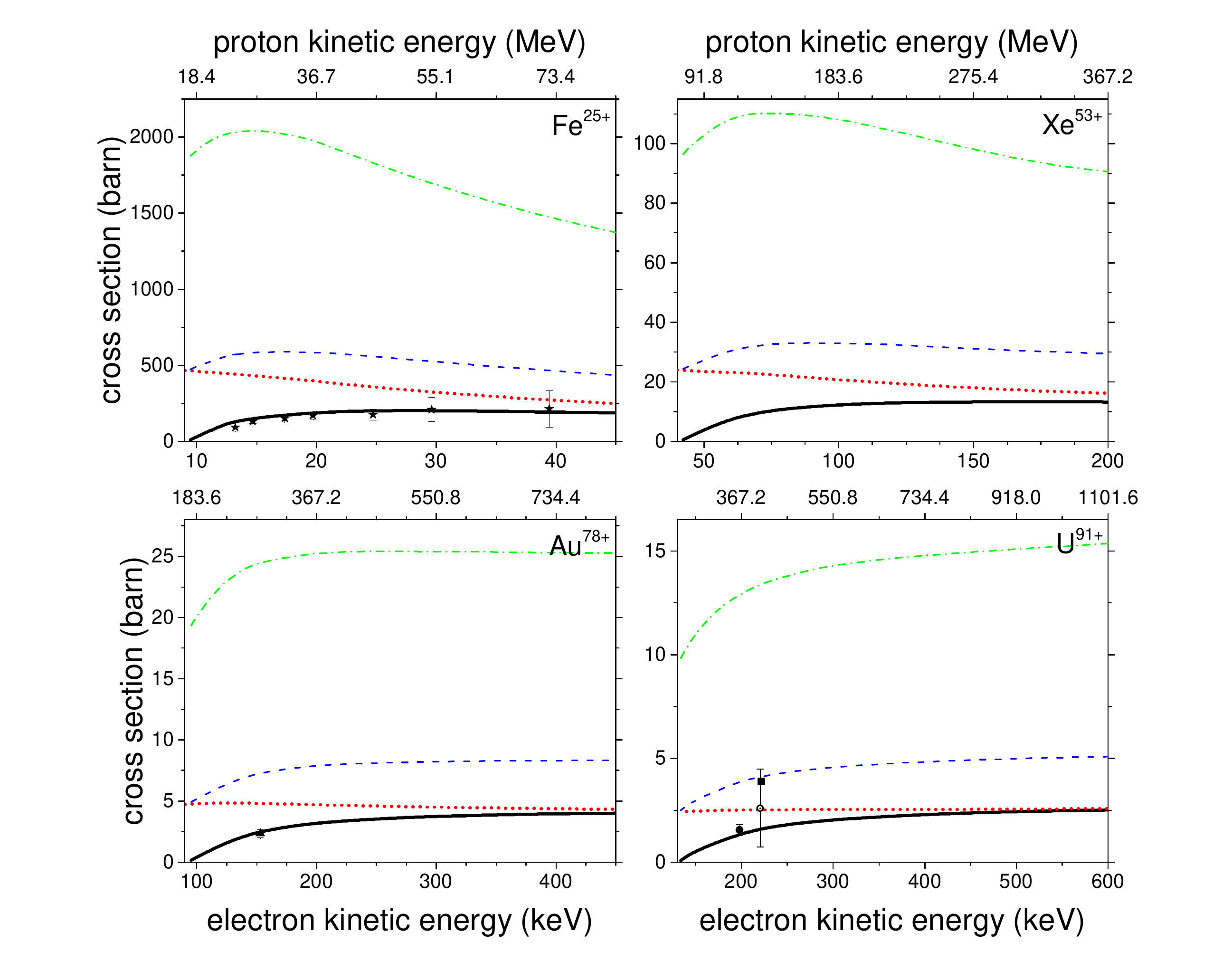}
\caption{The total cross sections for electron loss from Fe$^{25+}$(1s), 
Xe$^{53+}$(1s), Au$^{78+}$(1s) and U$^{91+}$(1s) in collisions with equivelocity electrons (solid curves), protons (dotted curves) 
as well as with atoms of hydrogen (dashed curves) and helium (dash-dot curves). 
In the right lower panel the open circle with error bars 
is the total cross section for electron loss from $405$ MeV/u U$^{90}$(1s$^2$) colliding with H$_2$ measured in \cite{hulskotter1991} (which we scaled to collisions of hydrogen-like uranium with atomic hydrogen by diving their result by $4$). All the other experimental data, shown in this figure, were measured for electron-ion collisions and are taken from \cite{rourke2001} (Fe$^{25+}$(1s), $\bigstar$), \cite{marrs1997} (Au$^{78+}$(1s), $\blacktriangle$), \cite{claytor1988} (U$^{91+}$(1s), 
$\blacksquare$) and \cite{marrs1994} (U$^{91+}$(1s), $\bullet$).}
\label{fig4}
\end{minipage}\hspace{2pc}%
\end{figure}

In Fig. \ref{fig4} we present results for the total cross section of electron loss from HCIs in collisions with equivelocity electrons, protons, atomic hydrogen (for collisions with randomly oriented hydrogen molecules these results should be multiplied by $2$) and helium. Four different HCIs are considered: they range from Fe$^{25+}(1s)$ till U$^{91+}(1s)$ covering, thus, quite a broad range of the atomic numbers of HCIs. 
For each HCI the total cross sections are given as a function of the electron impact energy (and the corresponding impact energy of an equivelocity proton). 
In each case the electron impact energy varies from the corresponding threshold 
value $\varepsilon_{th}$ to  about $5\varepsilon_{th}$. 

\par
The following main observations can be drawn from the results shown in Fig. \ref{fig4}. 
First, at impact velocities corresponding to the near-threshold values of the energy of the incident electron the protons are much more effective in producing electron loss than the equivelocity electrons. Second, with increasing the impact velocity the relative effectiveness of electron projectiles increases reaching almost "parity" with that of equivelocity protons at the highest velocities shown in the figure. Third, 
an interesting feature of electron loss in collisions with equivelocity electrons and protons is that the relative effectiveness of electrons increases with the increase of the atomic number of the HCI. 
Besides, comparing results for electron loss from uranium in Fig. \ref{fig4} and Figs. \ref{fig2}-\ref{fig2-add} we may conclude 
that the equality of the total loss cross sections in collisions 
with equivelocity electrons and protons by no means implies 
the same (or even similar) shape of the differential cross sections.  

It is of interest to compare the above observations with those which were made in 
\cite{najjari2013} where the process of excitation of HCIs into bound states by equivelocity electrons and protons was investigated. In particular, it was found there that:  
i) at electron impact energies near the excitation threshold the electrons 
are on average more effective in producing excitation than equivelocity protons, the cross section for excitation by electron impact reaches its maximum at the threshold; 
ii) with increase in the impact velocity the difference between equivelocity electrons and protons diminishes; and iii) the relative effectiveness of electrons in producing excitation increases when the atomic number of the HCI grows. 

Thus, we see that the processes of loss and excitation 
in collisions with electrons have both similarities and differences. 

The key difference is that the loss cross section is zero at the energy threshold for this reaction whereas the excitation cross section has a maximum 
at the excitation threshold. These qualitatively different behaviours of the cross sections are related to the fact that for the near-threshold collisions the smallness of the final-state phase space in case of excitation is fully compensated by the singularity in the Coulomb wave function of the slow outgoing electron whereas in case of electron loss, where there are two slow outgoing electrons whose energies are connected due to the energy conservation, such a compensation does not take place.    

The main similarity is that in both these processes 
the effectiveness of electrons compared to that of equivelocity protons 
grows with the increase in the atomic number of the HCI. Since according 
to the non-relativistic consideration the relative effectiveness must be independent 
of the HCI's atomic number (that can be very simply shown by scaling 
the Schr\"odinger equation to $Z_I$), this grows should be attributed 
to relativistic effects in the motion of the electrons and 
the interaction between them.   

In Fig. \ref{fig4} shown also are available experimental results on the total loss cross section 
from HCIs in collisions with electrons (to our knowledge, such data are absent 
for collisions with protons and atomic hydrogen). Besides, in this figure we also display the cross section for electron loss 
from $405$ MeV/u U$^{90+}$(1s$^2$) in collisions with H$_2$ measured in \cite{hulskotter1991} 
which we have divided by $4$ in order to scale it  
to collisions of $405$ MeV/u U$^{91+}$(1s) with atomic hydrogen.  

For electron loss from very heavy hydrogen-like HCIs the experimental 
data are very scarce (see Fig. \ref{fig4}) and do not seem to be very accurate. Besides, in case of electron loss from U$^{91+}(1s)$ 
by electron impact experimental data of different groups are not 
in agreement with each other.    

\section{ Conclusion }

We have compared the processes of electron loss from highly charged hydrogen-like ions 
by the impact of equivelocity electrons and protons 
in the range of impact velocities 
$v_{min} < v \leq v_{max}$, where 
$v_{min}$ and $v_{max}$ correspond to the energy threshold $\varepsilon_{th}$ 
for this reaction in collisions with electrons 
and to $\approx 5 \, \varepsilon_{th}$, respectively. 
We have also applied our results obtained for equivelocity 
electron and proton impact to the process of electron loss in collisions 
with hydrogen and helium. 

In the process of excitation by electron impact near the threshold 
for this reaction the smallness of the momentum space of the outgoing electron 
is fully compensated by the Coulomb singularity in its wave function \cite{najjari2013}.  
In contrast, in the process of electron loss near the threshold 
the smallness of the momentum spaces of the two outgoing electrons 
is not compensated by the Coulomb singularity in their wave functions. 
Thus, unlike the process of excitation, the Coulomb attraction between 
the electrons and the nucleus of the HCI cannot "compete" with the advantage 
of the very large phase (momentum) space available in collisions with protons 
and, therefore, the latter turn out to be more effective, 
compared to equivelocity electrons, in producing electron loss. 
However, similar to excitation, the relative effectiveness 
of electron projectiles increases when the atomic number of the HCI grows. 
This increase is a purely relativistic effect.      

In collisions with protons the spectra of outgoing electrons extend to much larger energies. Besides, by analysing the electron energy-angular distribution 
we found out that a substantial part of the final-state-electron momentum space, kinematically available in collisions with electrons, 
is stronger populated in collisions with electrons 
than in collisions with equivelocity protons. Taking all this into account 
we may conclude that in collisions 
of HCIs with hydrogen target the contributions to the electron loss process, 
which are caused by the interactions of the electron of 
the HCI with the nucleus(nuclei) and the electron(s) 
of the target, are to a large extent separated from each other 
in the energy-angular distribution 
and, by picking up 'appropriate pieces' of the momentum space, 
could be explored in experiment independently.  

It also follows from our results that even when, with increase in the impact energy,  
the total loss cross sections in collisions with electrons and protons 
become already equal, 
the spectra of the outgoing electrons still remain substantially different 
in almost the entire volume of the final-state-electron momentum space.  

In case of electron loss from very heavy hydrogen-like HCI (like e.g. gold, uranium) 
by electron impact experimental data are quite scarce even for the total cross section  
(sometimes contradicting to each other). Besides, we are not aware about any experimental results 
on electron spectra in such collisions. Further, accurate 
experimental results for the total cross section of electron loss from 
heavy hydrogen-like HCIs in collisions with very light targets are absent. 
To our knowledge, there is also no experimental data on the electron loss 
spectra in such collisions. 
All this clearly suggests that new experiments in this field are very desirable  
and the present theoretical results (as well as the availability 
of an accurate method for calculating electrons loss from HCIs in collisions 
with electrons, protons and very light atoms) could become a guide 
for further experimental activities in this field, in particular, 
for experiments on 
ion-atom collisions planned at the GSI (Darmstadt, Germany) and 
in the Institute of Modern Physics (Lanzhou, China).  
 
\section*{Acknowledgement} 

The work was partly supported by the DFG grant VO 1278/4-1. 
The work of K.N.L. and O.Yu.A. presented 
in section II.A. and III (collisions with electrons) 
was supported solely by the Russian Science Foundation under the grant 17-12-10135.
The work of K.N.L. was also partly supported by RFBR Grant 16-32-00620.
K.N.L. and O.Yu.A. acknowledge the hospitality of the Institute for 
Theoretical Physics I during their visit to the Heirich-Heine-University of 
D\"usseldorf.


\begin{thebibliography}{99}
\bibitem{foot-el-comp} The electron part of 
(\ref{atom}) should in general be averaged over 
the atomic Compton profile.
\bibitem{foot-wannier} The limitations of the first order approach  
in collisions with electrons is discussed in subsection C of section II.
\bibitem{foot-proton-cusp} The limitations of 
the first order approach in collisions with protons 
is discussed 
in subsection C of section II.
\bibitem{foot1} When calculating the total cross section the integration 
over the angles can be done analytically that makes the computation 
much easier and greatly saves the computation time.
\bibitem{foot-compton-profile} As was mentioned in \cite{foot-el-comp}, 
the contribution of the atomic electron to the loss cross sections should, in principle, 
be averaged over the atomic Compton profile. However, 
the Compton profile of hydrogen is quite narrow and 
the averaging would make a substantial difference only at impact velocities 
very close to the electron threshold where our approach, 
because of the reasons discussed in subsection C of section II, 
is no longer accurate.
\end{thebibliography}
\end{document}